# An exactly solvable model of random site-specific recombinations

Yi Wei and Alexei A. Koulakov
Cold Spring Harbor Laboratory, Cold Spring Harbor, New York 11724, USA

**Abstract**

Cre-lox and other systems are used as genetic tools to control site-specific recombination (SSR) events in genomic DNA. If multiple recombination sites are organized in a compact cluster within the same genome, a series of random recombination events may generate substantial cell specific genomic diversity. This diversity is used, for example, to distinguish neurons in the brain of the same multicellular mosaic organism, within the brainbow approach to neuronal connectome. In this paper we study an exactly solvable statistical model for SSR operating on a cluster of recombination sites. We consider two types of recombination events: inversions and excisions. Both of these events are available in the Cre-lox system. We derive three properties of the sequences generated by multiple recombination events. First, we describe the set of sequences that can in principle be generated by multiple inversions operating on the given initial sequence. We call this description the ergodicity theorem. On the basis of this description we calculate the number of sequences that can be generated from an initial sequence. This number of sequences is experimentally testable. Second, we demonstrate that after a large number of random inversions every sequence that can be generated is generated with equal probability. Lastly, we derive the equations for the probability to find a sequence as a function of time in the limit when excisions are much less frequent than inversions, such as in shufflon sequences.

## 1. Introduction

Site-specific DNA recombination is a useful mechanism that can generate genomic diversity. Mosaic animals, that carry distinguishable genomes in different cells of the same organism, have become important tools in studying its organization. For example, in brainbow mice, neurons carry randomly colored fluorescent dies, which allow to distinguish neighboring cells with the purpose to establish their connectivity (*1-3*). Next generation sequencing methods allow to trace stem cell lineage by identifying progeny with similar genetic barcodes (*4*). To implement this strategy, stem cells have to carry distinguishable genetic sequences. Site-specific DNA recombination could be used to generate such a sequence diversity. Finally, mosaic sequence diversity may help study connections between neurons using the next generation sequencing technologies (*5, 6*). Here we study the statistical model for SSR events with the purpose of understanding both the diversity of their products and their probability distributions.

Cre-lox system can be used as a tool to both control gene activation (*7-9*) and generate randomly diverse DNA sequences (*1-3*). A loxP site (locus of X-over P1) is a 34bp sequence segment consisting of two 13bp inverted complementary repeats separated by an oriented 8 bp asymmetric region (ATAACTTCGTATA – GCATACAT - TATACGAAGTTAT) (*9*). The orientation of loxP recombination site is determined by the central 8bp region. The Cre recombinase is an enzyme that mediates SSR at loxP sites. Such recombination events occur between two loxP sites and the outcome depends on the relative orientation of the loxP sites. When two loxP sites that have opposite orientation are recombined, an *inversion* occurs. After an inversion, the DNA segment between two loxPs is flipped: its orientation is inverted and the sequence is replaced with the complement (Figure 1) (*10*). The remaining parts of the original sequence which are not involved in the inversion are left unchanged. The recombination between two loxP sites of the same orientation yields an *excision*, whereby one of the loxPs and the whole segment between the two loxPs are removed from the sequence (Figure 2). Other variants of loxP have been identified, such as lox2272 and loxN, that can mediate Cre-based recombination to produce multicolored mouse brains with several fluorescent proteins in brainbow mice (*1-3*).

In other systems, such as rci-R64 recombination, the excision events are not reported (*11*). Instead, the recombination leads to an inversion. These systems are therefore called shufflons (*11*). For the most part, our study will be relevant to this type of system, because we will assume that excision events are either non-existent (Section 2 and 3) or rare (Section 4). In our study site specific recombination sites are called SSR-sites. Our model may apply to Cre-lox and other systems, such as rci-R64, Cre-lox2272/loxN, and others (*1-3, 11*).

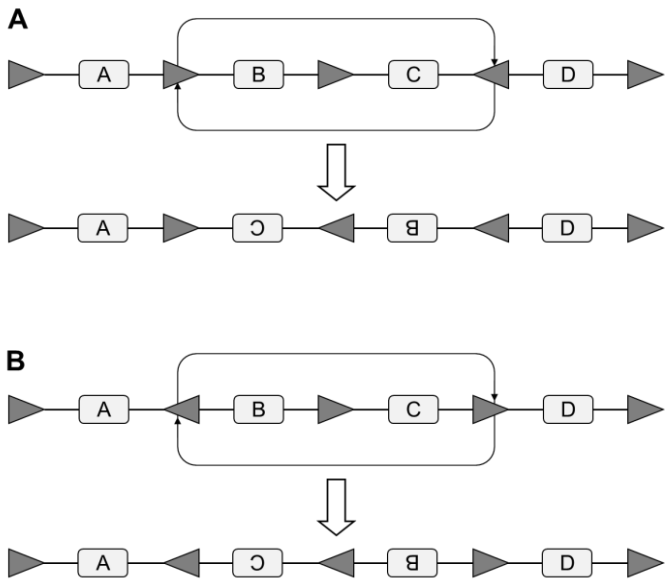

**Figure 1.** Two types of inversions. Inversion (A) is between an R (right) and an L (left) SSR-sites. Inversion (B) is between an L and an R SSR-sites.

We make several assumptions about the random recombination processes. First, we will assume that inversions and excisions can be described by independent Markov processes with probabilities that do not depend on time. Second, we will assume that when an event (inversion or excision) happens to a DNA sequence, it happens to all the SSR-sites, which satisfy the conditions for the corresponding recombination, independently with equal probability.

In this paper we answer the following questions regarding the inversion process:

**Q1.** Given an arbitrary initial DNA sequence, what are the sequences that can appear after applying an arbitrary number of inversions? What is the number of sequences that can be generated by such a process? These questions are addressed in Section 2 below.

**Q2.** Is there a unique equilibrium distribution of the DNA sequence configurations after a sufficiently large number of random inversions? What is the probability of observing one particular DNA sequence when the equilibrium is reached after a large number of inversions? This question is addressed in Section 3.

Combining both inversion and excision processes, we will answer to the following question.

**Q3.** Assume that excisions are much slower than inversion, so that excisions occur after an equilibrium due to inversions has been reached. Given an arbitrary initial DNA sequence at time T=0, what is the probability of observing a specific DNA sequence configuration at time T=t? This question is addressed in Section 4 of this study.

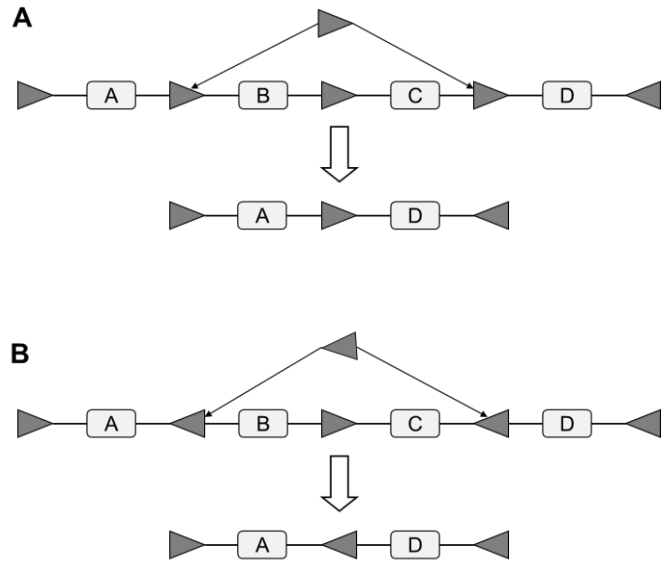

**Figure 2.** Two examples of excisions. Excision (A) is between two R SSR-sites and excision (B) is between two L SSR-sites.

## 2. The diversity of the results of multiple inversions

In this section we will derive the sequences that can be obtained from an initial DNA sequence after a number of inversions. We will study the cluster of SSR-sites of different orientations located on the same DNA segment in close proximity so that inversions are possible between any pair of appropriate orientation. We will assume that inter SSR-sites DNA segments carry distinguishable sequences. We will enumerate the possible sequences that can be produced by multiple inversion events, given this initial sequence.

Before we rigorously derive our result, we briefly present our main finding of this Section. Given an initial sequence (Figure 3A), all of the sequences that can be generated by inversions can be obtained as follows. First, one cuts the sequence into segments containing nearest pairs of SSR-sites and inter-SSR-sites DNA sequences. We call such segments units. Second, one builds a dictionary of units that contains both direct and inverted sequences (Figure 3B). These units can be recombined into new sequences (Figure 3C) by satisfying the following two rules. The first rule is that each unit from the original sequence should be used once and only once. The second rule postulates that the orientations of SSR-sites have to agree between the edges of neighboring units (Figure 3C). In this Section we show that any sequence that follows these two simple rules can be obtained by inversions from the original sequence. This means that inversions can generate any sequence within a cluster of SSR-sites that does not generate new units. We call this property of the inversion process, proven in Theorem 2, the ergodicity property.

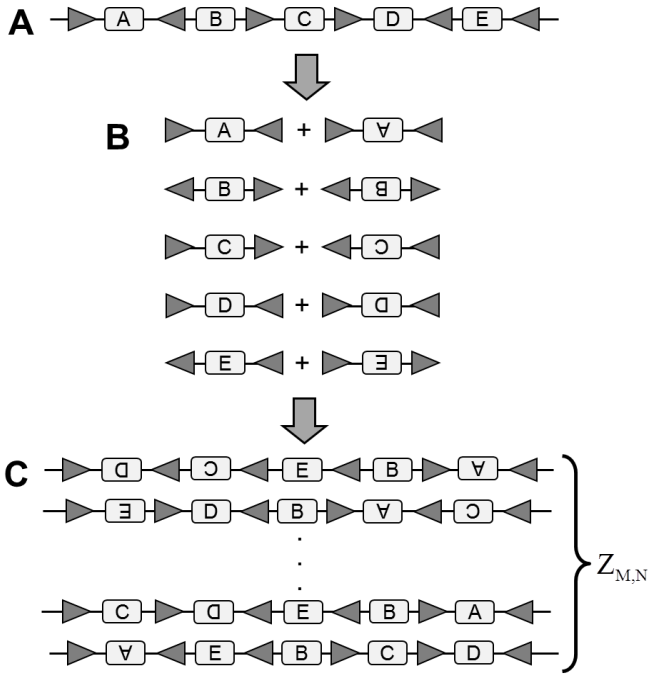

**Figure 3.** Ergodicity of inversions. To enumerate all sequences that can be obtained from an original sequence (A) by inversions, one first generates the dictionary of units (B). The units include direct and inverted segments located between two neighboring SSR-sites. Units also include the information about the direction of SSR-sites (gray triangles). The units can be recombined into new sequences that include each unit only once and respect the orientation of SSR-sites (C). All of the sequences that satisfy these constraints can be produced by inversions from the original sequence (A). This ergodicity property of inversions is proven here in Theorem 2. The total number of such sequences is given by Eq. (2.2) and is denoted by $Z_{M,N}$.

In a DNA sequence with SSR-sites, every two adjacent SSR-sites and the DNA segment between them form a unit. Two adjacent units share one common SSR-site. A DNA sequence can be viewed as a chain of such units. It is convenient to classify units according to the orientations of their SSR-sites. For example, an RR unit and an RL unit are shown in Figure 4.

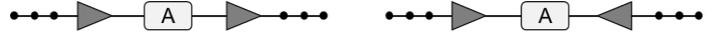

**Figure 4.** Examples of an RR unit and an RL unit.

In order to find all possible configurations resulting from inversions, it is important to note some properties of the inversion operation of DNA sequences.

**Property 1.** An RR unit is transformed into an LL unit if it is inside the part of DNA sequence that has been inverted (Figure 5). Otherwise it remains unchanged. Similarly, an LL unit will either be transformed into an RR unit or remain unchanged. An RL (LR) unit remains an RL (LR), but the DNA sequence between SSR-sites is either unchanged, if the unit is not inverted, or changed to reverse complement as a result of inversion.

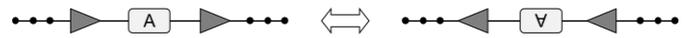

**Figure 5.** An RR unit transforms to an LL unit after an inversion.

**Property 2.** The orientations of SSR-sites on both ends of the whole DNA sequence remain invariant under all inversions and excisions.

**Property 3.** RL units and LR units are distributed alternatively along a sequence. An RL unit is separated by RR units (or nothing) from its left neighboring LR unit and by LL units (or nothing) from its right neighboring LR unit.

These observations lead to the following definition.

**Definition 1.** The rank N of a DNA sequence is the total number of its RL and LR units, i.e. $N = N_{RL} + N_{LR}$. By Property 1, N is invariant with respect to inversions.

The configuration of a DNA sequence, including the orders and the orientations of the SSR-sites and DNA segments, can be decomposed into the SSR-site part and the inter-SSR-site segment part. The SSR-site part defines a SSR-site array, see Figure 6. It is convenient to work first in the SSR-site space, defined as the set of all possible SSR-site configurations of a given DNA sequence, and to restore the segment part in the final step. For SSR-site array, a unit is a pair of adjacent SSR-sites (which is just a unit in the DNA sequence without the DNA segment).

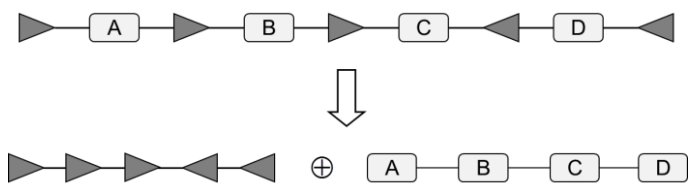

**Figure 6.** A configuration of a DNA sequence can be represented as a 'direct sum' of its SSR-site array and inter-SSR-site segment array.

Let us assume the DNA sequences start with an R SSR-site.

**Definition 2.** A SSR-site array is called in the canonical form if it has no LL units and all the RR units are located on the left part of the sequence (Figure 7).

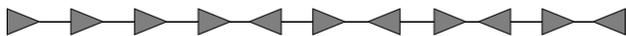

**Figure 7.** A SSR-site array in the canonical form, beginning with 3 RR units followed by alternating RL and LR units.

**Theorem 1.** Any SSR-site array can be transformed into the canonical form with a finite number of inversions.

Proof: For notational convenience, we assign fixed symbols to certain SSR-sites in an array. In the process of inversions, a SSR-site may change its orientation as well as position but its symbol is never changed and always attached to it. Denote an inversion between SSR-site a and b by I=(a, b). The composition of two consecutive inversions, first between a and b and then between c and d, is written as I=(a, b)∘(c, d). The same notation is adopted for multiple inversions.

Next we define two types of procedures, as shown in Figure 8. Note that after each procedure, the first two SSR-sites are changed from RL to RR.

1. As already mentioned, we always assume an array starts with an R SSR-site. If the array is partially in the canonical form, i.e. it begins with a continuous series of RR's and then a series of alternating RL and LR until an occurrence of an LL, perform a type-A inversion between this LL and the first RL, as shown in Figure 8A.

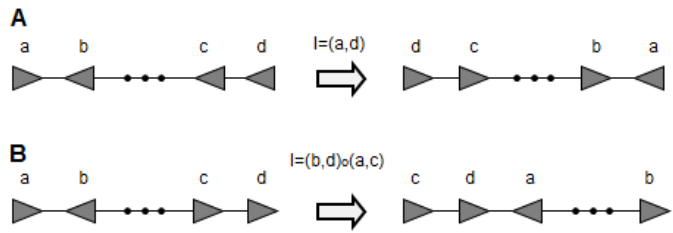

**Figure 8.** (A). Type-A inversion. (B). Type-B inversion.

2. If the sequence is partially in the canonical form until an occurrence of an RR to the right of the canonical part, perform a type-B inversion between this RR and the first RL, as shown in Figure 8B.

The result of each of these two operations is that the subsequence in the canonical form is increased by 1 unit. Repeat step 1-2, until there is no more LL or RR unit to the right of any L SSR-site, i.e. the whole sequence is in the canonical form. □

By Theorem 1, if a set of SSR-site arrays can be transformed into the same canonical form, they can be transformed into each other with a finite number of inversions.

**Note:** Without loss of generality, we will only study DNA sequences starting with an R and ending with an L SSR-site in remaining part of this paper. Hence, all formulae in the remaining of this paper are for DNA sequences of this particular form. Results for sequences ending with an R SSR-site can be derived in the same way and are essentially the same.

**Definition 3.** If a SSR-site array A can be transformed into another array B by a finite number of inversions, we say A and B are equivalent, denoted by A ~ B.

Clearly, if A ~ B and B ~ C, then A ~ C.

By Property 3, the number of RL and LR units of an array are both invariant with respect to inversions. In fact, it is easy to show that a rank-N SSR-site array has $[\frac{N+1}{2}]$ number of RL and $[\frac{N}{2}]$ number of LR, where $[x] = \max(n \leq x \mid n \text{ is integer})$. Let (M, N) denote the set of rank-N SSR-site arrays with M number of RR and LL units, i.e. $M = \sharp RR + \sharp LL$. By Theorem 1, all arrays in (M, N) can be transformed into the same canonical form, which is just a special element in the set which has M number of RR units. Therefore any two arrays in (M, N) can be transformed into each other by a finite number of inversions, i.e. arrays in (M, N) are equivalent to each other. This observation leads to the following conclusion.

**Property 4.** The set (M, N) of SSR-site arrays forms an equivalence class.

**Definition 4.** Let F be a function defined on SSR-site arrays with fixed length. If the value of F is determined only by the rank of sequences, F is called a class function. A class function takes a constant value in an equivalence class (M, N).

By Property 3 and Definition 1, arrays with different ranks are not equivalent. Let S be the set of SSR-site arrays with N total units. S is decomposed into a direct sum of N number of equivalence classes with different ranks, i.e. S=(N-1,1) $\oplus \cdots \oplus$ (0,N). Under all possible inversions, a SSR-site array will span the whole equivalence class it belongs to.

We are now ready to restore the DNA segments and answer **Q1** based on the properties of SSR-site arrays. First, we make the following observation.

**Property 5.** If the SSR-site array of a DNA sequence is in the canonical form, all permutations of its RR units can be achieved by inversions, while keeping the array in the canonical form and configurations of RL and LR units unchanged. Similarly, all permutations of the RL (LR) units can be achieved without affecting the RR and LR (RL) parts.

Proof: 1. We first show that an arbitrary pair of segments of neighboring RR units can be exchanged while leaving the remaining array unchanged. It is sufficient to elucidate the procedure with an example shown in Figure 9. It is straightforward to check that the operation $I_1 = (a,d) \circ (b,d) \circ (a,c) \circ (c,d)$ exchanges segments A and B and leaves all other units unchanged. Since an arbitrary permutation of segments in RR units can be generated by a series of pair exchanges (because very element of symmetry group $S_N$ can be written as a product of exchanges), we complete the proof of the first claim.

2. Similarly, in Figure 9, $I_2 = (d,g) \circ (d,e) \circ (d,f) \circ (d,g)$ exchanges segments D and F and leaves all other units unchanged. By the same argument above, we conclude that when the SSR-site array is in the canonical form, we can get, by inversions, all permutations of segments of LR units without modifying the remaining of the sequence. In the same way we can prove the same argument holds for RL units. □

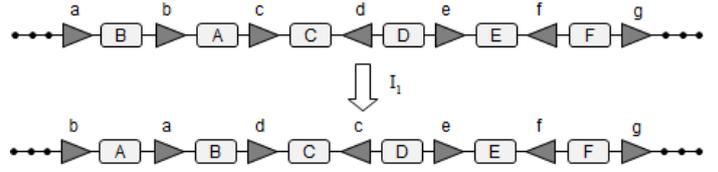

**Figure 9.** An example of exchanging two RR units without modifying all others units in the sequence.

Using Theorem 1 and Property 1-5, we now summarize this section by answering **Q1** with the following theorem.

**Theorem 2.** (Ergodicity) Let the SSR-site configuration of the initial DNA sequence be a state in (M, N). Any DNA sequence satisfying the following three conditions can be reached by applying a finite number of inversions on the initial sequence. C1. Its SSR-site configuration is a state in the same class (M, N); C2. Its RL (LR) units, or their inversions, are the RL (LR) units of the initial sequence; C3. Its RR (LL) units are the RR (LL) units, or LL (RR) units with inverted inter-SSR-site segments, of the initial sequence.

The total number of possible SSR-site arrays in class (M,N) is given by

$$d_{(M,N)} = C_{M+N}^{N} = \frac{(N+M)!}{N!M!}. \quad (2.1)$$

For a given initial sequence, assuming that inter-SSR-site sequences are distinguishable and non-symmetric, the total number of unique configurations that can be obtained by inversions is

$$Z_{M,N} = 2^N M! \left[\frac{N+1}{2}\right]! \left[\frac{N}{2}\right]! d_{(M,N)}. \quad (2.2)$$

Where, as before, we used the notation $[x] = \max(n \le x \mid n \text{ is integer})$. According to Theorem 2, all of these sequences can be reached by a finite set of inversions from any initial sequence. The number of configurations $Z_{M,N}$ can be measured using DNA sequencing and is therefore an experimentally testable prediction.

## 3. Probabilities and statistics associated with random inversions

In this section we study the statistical properties of SSRs under the random inversions. By our initial assumption, all possible inversions of any sequence are independent and happen with the same probability. Under this assumption, here we show that after a sufficiently large number of inversions, the configurations reach a unique equilibrium distribution. We show that in this distribution all possible configurations described in the previous chapter are equally likely (Theorem 3).

We first define the random process that will be used in this chapter. We will denote the probability to observe a configuration number $i$ at time $t$ by the vector $\Psi_i(t)$. The configuration may be defined either by the array of SSR-sites of or by both SSR-sites and intermediate DNA sequences. We will derive general properties of the random process first without making the definition of the configuration more concrete. Because all inversions occur with the same probability, we can relate two vectors at two near time points separated by a short interval $\Delta t$:

$$\Psi_i(t+\Delta t)=r_{inv}\Delta t\sum_j \hat{R}_{ij}\Psi_j(t)+\Psi_i(t)\left[1-r_{inv}\Delta t\sum_k \hat{R}_{ki}\right] \quad (3.1)$$

In this expression $r_{inv}$ is the rate with which a single inversion occurs, $\hat{R}_{ij}$ is the number of inversion that connects states $i$ and $j$ that is usually either 0 or 1, and the last term is needed to ensure the conservation of probability, i.e. that $\sum_i \Psi_i(t)=1$ for any $t$. By differentiating this expression with respect to $\Delta t$ and setting it to 0, we obtain

$$\frac{d\Psi_i(t)}{dt}=\sum_j r_{inv}\cdot\left(\hat{R}_{ij}-C_i\delta_{ij}\right)\Psi_j(t). \quad (3.2)$$

Here $C_i=\sum_j \hat{R}_{ji}$. This equation has the following solution

$$\vec{\Psi}(t)=\exp\left(r_{inv}\left[\hat{R}-\hat{C}\right]t\right)\vec{\Psi}(0), \quad (3.3)$$

Where the elements of the diagonal matrix are given by $C_{ij}=C_i\delta_{ij}$. Using this equation we will study the equilibrium distribution of sequences at the end of a large number of rotations.

Equation (3.3) describes a continuous-time Markov process with the transition probability matrix $\hat{T}\equiv\exp\left(r_{inv}\left[\hat{R}-\hat{C}\right]t\right)$. The equilibrium distribution of this process can be obtained as a limit $\vec{\Psi}=\lim_{t\to\infty}\exp\left(r_{inv}\left[\hat{R}-\hat{C}\right]t\right)\vec{\Psi}(0)$. Because of the ergodicity theorem 2 and the fact that $\hat{T}$ is non-negative, this distribution is unique and does not depend on the initial state $\vec{\Psi}(0)$. Clearly, the elements of $\Psi_k$ are non-zero only for a subset of states that are reachable from the initial configuration, i.e. for $Z_{M,N}$ of such states [Eq. (2.2)]. Because the transition matrix $\hat{T}$ is symmetric, the non-zero elements of $\Psi_k$ are the same and equal to $1/Z_{M,N}$ (*12*). This follows from the detailed balance condition pertinent to symmetric Markov processes in the equilibrium, i.e. $T_{kn}\Psi_n=T_{nk}\Psi_k$ (*12*). We therefore arrive to the following theorem:

**Theorem 3:** There exists a unique vector $\Psi_\infty$, such that for an arbitrary initial state $\Psi(0)$, $\lim_{t\to\infty}\hat{T}(t)\Psi(0)=\Psi_\infty$. All non-zero components of vector $\Psi_\infty$ are the same.

This theorem means that all of the states that can be reached from an arbitrary initial state are represented in the equilibrium with equal probability. Interestingly, this result is valid for both configurations that include SSR-sites only and the complete sequences.

To give an example of the transition matrices we will consider the sequence defined by SSR-site configuration only (Figure 10). The possible states of recombination sites only (Figure 10, bottom) help define the transition matrix $\hat{R}=\begin{pmatrix}1 & 1 & 1\\ 1 & 2 & 1\\ 1 & 1 & 1\end{pmatrix}$. This matrix means that there are two inversions that leave state 2 invariant and there is one inversion for all other transitions. The rate of transitions is determined by the matrix $\hat{R}-\hat{C}=\begin{pmatrix}-2 & 1 & 1\\ 1 & -2 & 1\\ 1 & 1 & -2\end{pmatrix}$ that clearly has an eigenvalue of $\lambda=0$ corresponding to the constant eigenvector $\Psi_\infty=\begin{pmatrix}1/3 & 1/3 & 1/3\end{pmatrix}$. This eigenvector represents the distribution at $t\to\infty$. The other eigenvalue $\lambda=-3$ corresponds to the eigenvectors that decays over time.

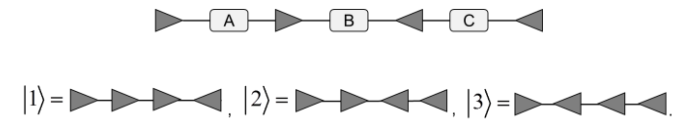

**Figure 10**. An example of SSR-site-only configuration. Top: The initial sequence. Bottom: Possible configurations of SSR-sites.

If the transition matrix is defined for the complete sequences, we anticipate that the elements of $\hat{R}$ can be either 0 or 1. This is a consequence of the inter SSR-sites

DNA segment having unique sequences. The matrix $_r\hat{R}$ defines the transition rates for the inversion processes. Our Proposition 1 relied on the symmetry of this matrix only. Thus, if inversion rates depend on the distance within the pair of SSR-sites, as long as the inversion process is reversible, Theorem 3 is expected to be valid. Thus, we suggest that the equal probability Theorem 3 is valid even if not all of the inversions are equally likely and transition probabilities are dependent upon the length of inter SSR-sites stretch.

## 4. Probabilities and statistics of inversions and excisions.

In this section we consider both inversions and excisions. We will answer **Q3** of the introduction, i.e. we will derive the probability of observing a sequence given the initial sequence as a function of time. We will assume that the excisions events are much rarer than the inversion events. We will assume that between two excisions, the sequence distribution reaches equilibrium due to inversions. This assumption allows us to find solutions for the probability of observing a sequence. We will find the answer by two different methods. First, we will use the summation over paths to determine probabilities of transitions. Second, we will employ the method based on matrix exponential. We will show that these methods give the same result.

First we will consider a simplified version of **Q3** of the Introduction. We will only address the SSR-sites and will not include inter-SSR-site DNA segments. The probability distribution of the latter can be evaluated on the basis of equal probability argument (Theorem 3).

**Q3':** Starting with a state in a class $(M_0, N_0)$ at time T=0, what is the probability of finding the sequence in the class $(M_1, N_1)$ at T=t?

First we will use the summation over paths method. To answer Q3', we note that all possible $(M_1, N_1)$ which are reachable from $(M_0, N_0)$ must satisfy the conditions $\Delta M = M_0 - M_1 \geq 0$ and $\Delta N = N_0 - N_1 = 2k$, where $k \geq 0$. In fact, all classes $(M_1, N_1)$ that satisfy these two conditions can be reached from certain states of $(M_0, N_0)$ with a single (1-step) excision. Although it is not necessarily true that the class $(M_1, N_1)$ can be reached by 1-step excision from every state of $(M_0, N_0)$.

We define two quantities associated with the excision transitions between two classes. Let $x_i^{(M,N)}$ be the total number of possible excisions of the i-th state of the class $(M, N)$. Then we have $x_i^{(M,N)} = W - C_i$, where $W = \frac{1}{2}(M+N)(M+N+1)$ and $C_i$ is the connectivity of the state i of class $(M, N)$, defined as $C_i = (r + [\frac{N+1}{2}])(M - r + [\frac{N+1}{2}])$. Here r is the number of RR units. Define $x_{(M,N)}$ as the average number of excisions of states in $(M, N)$, then we have $x_{(M,N)} = W - \frac{1}{d_{(M,N)}} \sum_{i \in (M,N)} C_i$, where $d_{(M,N)}$ is the total number of states in the class $(M, N)$ [(Eq. (2.1)]. Define $D_{(M,N)}(r)$ as the number of states in a class $(M, N)$ which has r number of RR units. It is straightforward to show that $D_{(M,N)}(r) = C_{r+[\frac{N}{2}]}^{r} \times C_{M-r+[\frac{N-1}{2}]}^{M-r}$. Then we have

$$\sum_{i \in (M,N)} C_i = \sum_{r=0}^{M} D_{(M,N)}(r) C(r)$$

$$= \begin{cases} k^2 C_{M+2k+1}^M, & N = 2k-1 \\ k(k-1) C_{M+2k}^M, & N = 2k-2 \end{cases}.$$

Let $y_i^{(M_0,N_0) \to (M_1,N_1)}$ be the total number of 1-step excisions leading to $(M_1, N_1)$ from the i-th state of $(M_0, N_0)$ and $\bar{y}^{(M_0,N_0) \to (M_1,N_1)}$ be the average of $y_i^{(M_0,N_0) \to (M_1,N_1)}$, i.e. $\bar{y}^{(M_0,N_0) \to (M_1,N_1)} = \frac{1}{d_{(M_0,N_0)}} \sum_{i \in (M_0,N_0)} y_i^{(M_0,N_0) \to (M_1,N_1)}$. The sum is calculated by noticing the following facts. For a state $\mu \in (M_1, N_1)$, let $\bar{\mu}$ be the set of SSR-site array which, when inserted into μ as in Figure 11, make the combined arrays states of $(M_0, N_0)$. There are two types of arrays in the set $\bar{\mu}$, see Figure 11. Type-I arrays begin and end with R SSR-site and correspond to the excisions shown in Figure 2A. By definition, type-I sequences belong to the class $(\Delta M, \Delta N)$, where $\Delta M = M_0 - M_1$ and $\Delta N = N_0 - N_1$. Similarly, type-II arrays belong to the equivalence class $(\Delta M, \Delta N)^T$, which begin and end with L SSR-sites, corresponding to the excisions shown in Figure 2B.

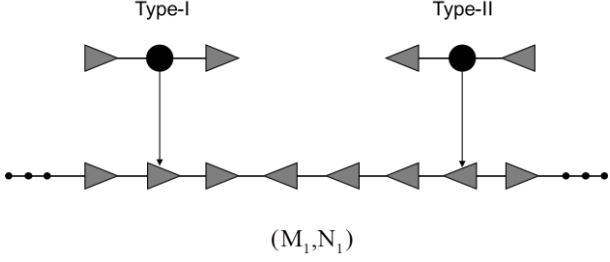

**Figure 11**. Type-I arrays belong to the class $(\Delta M, \Delta N)$. A type-I arrays can be inserted to each R SSR-site of a state in $(M_1, N_1)$, and the outcome is an array in $(M_0, N_0)$. Type-II arrays belong to the class $(\Delta M, \Delta N)^T$ and should be inserted to L SSR-sites.

There is a 1-to-1 correspondence between the arrays of type-I and type-II by inversion. Let $v_I$ be an array of type-I, then the inversion of $v_I$ is an array of type-II and is denoted by $v_I^T$. From Figure 11 caption we find

$$\bar{y}^{(M_0,N_0) \to (M_1,N_1)} = \frac{1}{d_{(M_0,N_0)}} \sum_{i \in (M_0,N_0)} y_i^{(M_0,N_0) \to (M_1,N_1)}$$
$$= (M_1 + N_1 + 1) \frac{d_{(M_1,N_1)} d_{(M_0-M_1, N_0-N_1)}}{d_{(M_0,N_0)}}. \quad (4.1)$$

Note that although type-I and -II arrays are not the typical SSR-site arrays we have being working with, i.e. begin with an R site and end with an L site, the dimension formula [Eq. (2.1)] still applies. This is because they both can be transformed into corresponding canonical forms and then all similar derivations follow.

Let $P^{k\text{-step}}_{(M_0,N_0) \to (M_1,N_1)}(t)$ be the probability of starting with a state in $(M_0, N_0)$ at T=0 and finding the array in a state of $(M_1,N_1)$ at T=t after k excisions. Denote the total probability by $P_{(M_0,N_0) \to (M_1,N_1)}(t)$, i.e. $P_{(M_0,N_0) \to (M_1,N_1)}(t) = \sum_{k=0}^{k_{max}} P^{k\text{-step}}_{(M_0,N_0) \to (M_1,N_1)}(t)$, where $k_{max} = \Delta M + [\frac{\Delta N}{2}]$. Assume that the rate of inversions $r_{inv}$ is much higher than the rate of excisions $r_{exc}$, i.e. $r_{inv} \gg r_{exc}$, so that inversions reach equilibrium between excisions. Then the probability to remain within class $(M_0, N_0)$ is given by $P_{(M_0,N_0) \to (M_0,N_0)}(t) = \exp(-r_{exc} x_0 t)$, where we used a shorthand notation $x_0 = x_{(M_0,N_0)}$. The probability of transition from $(M_0, N_0)$ to $(M_1, N_1)$ by time t with one excision is

$$P^{1\text{-step}}_{(M_0,N_0) \to (M_1,N_1)}(t) = \int_0^t dt'\, y_{01} r_{exc} \exp[-r_{exc} x_0 t' - r_{exc} x_1 (t-t')]$$
$$= \frac{y_{01}}{x_0 - x_1}[\exp(-r_{exc} x_1 t) - \exp(-r_{exc} x_0 t)]$$
(4.2)

where $y_{01} = \bar{y}^{(M_0,N_0) \to (M_1,N_1)}$ and $x_1 = x_{(M_1,N_1)}$. Similarly, the k-step transition probability through an excision path $\sigma: (M_0, N_0) \to (M_1, N_1) \to \cdots \to (M_k, N_k)$, can be computed as

$$P_\sigma(t) = \int_0^t dt_0\, y_{0,1} r_{exc} \exp(-r_{exc} x_0 t_0) \int_{t_0}^t dt_1\, y_{1,2} r_{exc} \exp[-r_{exc} x_1 (t_1 - t_0)] \ldots$$
$$\ldots \int_{t_{k-2}}^t dt_{k-1}\, y_{k-1,k} r_{exc} \exp[-r_{exc} x_{k-1}(t_{k-1}-t_{k-2}) - r_{exc} x_k (t-t_{k-1})]$$
$$= \left(\prod_{i=0}^{k-1} y_{i,i+1}\right) \frac{\det M(t)}{\Delta(x_0, \cdots, x_k)}$$
(4.3)

where $\Delta(x_0, \cdots, x_k) = \prod_{0 \le i < j \le k}(x_j - x_i)$ is the determinant of the Vandermonde matrix of $\vec{x}$ (*13*) and we defined

$$M(t) = \begin{pmatrix} \exp(-\lambda_{exc} x_0 t) & \exp(-\lambda_{exc} x_1 t) & \cdots & \exp(-\lambda_{exc} x_k t) \\ 1 & 1 & \cdots & 1 \\ x_0 & x_1 & \cdots & x_k \\ \vdots & \vdots & \vdots & \vdots \\ x_0^{k-1} & x_1^{k-1} & \cdots & x_k^{k-1} \end{pmatrix}$$
(4.4)

Note that in the derivation of Eq.(4.2)-(4.3), we made the approximation by assuming that the array reaches the equilibrium of inversions right after each excision. This approximation is valid only when inversions happen very fast, i.e. $r_{inv} \gg r_{exc}$ or the number of excisions in [0, t] is small. This is the case by our assumption.

To find $P^{k\text{-step}}_{(M_0,N_0) \to (M_1,N_1)}(t)$, we need to sum over all possible excision paths,

$$P^{k\text{-step}}_{(M_0,N_0) \to (M_k,N_k)}(t) = \sum_\sigma P_\sigma(t), \quad (4.5)$$

where the sum is over all k-step transitions between $(M_0, N_0)$ and $(M_k, N_k)$.

To find $P_{(M_0,N_0) \to (M_1,N_1)}(t)$, we define two $M[N/2] \times M[N/2]$ dimensional matrices: 1. an upper triangular matrix

$Y=(Y_{ij})$, $i,j=1,...,M[N/2]$, with elements $Y_{ij}=\overline{y}^{(M_i,N_i)\to(M_j,N_j)}$, for $i<j$, and $Y_{ij}=0, i \geq j$. 2. a diagonal matrix $\hat{X}$, with elements $X_{ii}=x_{(M_i,N_i)}$, $i=1,...,M[N/2]$. According to the properties of $x_{(M,N)}$, the matrix $\hat{X}$ is such that $X_{ii}=\sum_j Y_{ij}$.

The following theorem answers **Q3'**:

**Theorem 4.** Let $P_{(M_i,N_i)\to(M_j,N_j)}(t)$ be the probability of starting from an array in $(M_i,N_i)$ at T=0 and finding the array belongs to $(M_j,N_j)$ at T=t. Then we have

$$P_{(M_i,N_i)\to(M_j,N_j)}(t)=\left(\exp\left[\lambda_{exc}t(\hat{Y}-\hat{X})\right]\right)_{ij}. \quad (4.6)$$

Proof: Define a matrix P(t), with $P_{ij}(t)=P_{(M_i,N_i)\to(M_j,N_j)}(t)$, $i,j=1,...,M[N/2]$. By definition, we have

$$P_{ij}(t+dt)=P_{ij}(t)(1-X_{jj}r_{exc}dt)+\sum_k P_{ik}(t)Y_{kj}r_{exc}dt. \quad (4.7)$$

From which, we find P(t) satisfies the matrix equation

$$\frac{d\hat{P}(t)}{dt}=r_{exc}\hat{P}(\hat{Y}-\hat{X}). \quad (4.8)$$

Solving this equation, we find $\hat{P}(t)=\exp\left[r_{exc}t(\hat{Y}-\hat{X})\right]$. □

To check that Theorem 4 agrees with Eq. (4.5), we collect terms in the Taylor expansion of Eq. (4.6) with k number of Y's and find coefficients corresponding to each transition path $\sigma$ equal to those in Eq. (4.4).

We can now restore the inter-SSR-site DNA segments. Let $\mathbb{P}(t)$ be the probability of starting with a DNA sequence in $(M_0,N_0)$ at T=0 and finding a particular DNA sequence in $(M_1,N_1)$ at T=t. By the symmetry among RR and LL units stated in Theorem 3, to have the desired set of RR and LL units in the final sequence, we need to multiply $P_{(M_0,N_0)\to(M_1,N_1)}(t)$ by a factor $1/C_{M_0}^{M_1}$. Similarly, we need to multiply $P_{(M_0,N_0)\to(M_1,N_1)}(t)$ by factors $1/C_{[\frac{N_0}{2}+1]}^{[\frac{N_1}{2}+1]}$ and $1/C_{[\frac{N_0}{2}]}^{[\frac{N_1}{2}]}$ to have the desired RL and LR units. And to have all the segments in the correct order and orientation, we need to multiply by the probability by $1/Z_{M_1,N_1}$ given by Eq. (2.2). Define the quantity $Z=C_{M_0}^{M_1}C_{[\frac{N_0}{2}+1]}^{[\frac{N_1}{2}+1]}C_{[\frac{N_0}{2}]}^{[\frac{N_1}{2}]}Z_{M_1,N_1}$, finally we get

$$\mathbb{P}(t)=\frac{1}{Z}P_{(M_0,N_0)\to(M_1,N_1)}(t). \quad (4.9)$$

Theorem 4, Eq.(4.5) and Eq.(4.9) answer **Q3**.

## 5 Discussion

In this paper we derived the properties of random recombinations operating within a cluster of loxP or similar SSR-sites (*1-3, 11*). We addressed several questions pertaining to the properties of distributions of the resulting sequences. First, in Sections 2 and 3, we analyzed the processes of inversions only. We assumed that excisions between two recombination sites are nonexistent. This approximation is appropriate for the rci-R64 recombination system that is known as shufflon (*11*). Later, in Section 4, we included excisions into consideration assuming that they occur very infrequently.

We obtained two important results regarding the processes of inversion. First, we showed what DNA sequences are possible to obtain from an original sequence by an arbitrary set of inversions. The main conclusion is that any sequence that includes segments from the original sequence can be reached. Therefore we called this property the ergodicity of the inversion process. This property is defined more precisely in Theorem 2. On the basis of this property we derived the number of sequences that can be produced from any initial sequence by an arbitrary set of inversions. This number is given by equation (2.2) and is an experimentally testable prediction of our theory. This number of sequences can be studied using modern sequencing methods. In particular, assume that one initially has a large population of identical sequences of SSR-sites. Subsequent introduction of a restriction enzyme, such as rci or Cre, may lead to inversions that diversify the sequences. The number of resulting unique sequences can be counted by using sequencing methods and compared to Eq, (2.2). Below we present some examples of using this equation in simple cases.

Overall, our conclusions in Theorem 2 and 3 suggest that site specific DNA recombination can lead to diverse ensembles of sequences. Further, in Section 4, we derive an equation for the probability of obtaining sequences as a function of time when excisions are included, assuming that these processes are much slower than inversions.

**Simple example**

To illustrate our findings, let us consider the DNA sequence shown in Figure 10. Since the SSR-site array contains two RR or LL units and only one RL unit, this sequence belongs to the class (2,1), i.e. M=2, N=1. By Eq. (2.1), its SSR-site array has $d_{(2,1)} = C_{M+N}^N = C_3^1 = 3$ configurations, shown in Figure 10 above.

By Eq. (2.2), the total number of possible DNA sequences is $Z_{M,N} = 2^N M! \, [\frac{N+1}{2}]! \, [\frac{N}{2}]! \, d_{(M,N)} = 12$. Here and above the notation $[x]$ means the largest integer smaller than or equal to $x$. We show all these 12 configurations explicitly in Figure 12.

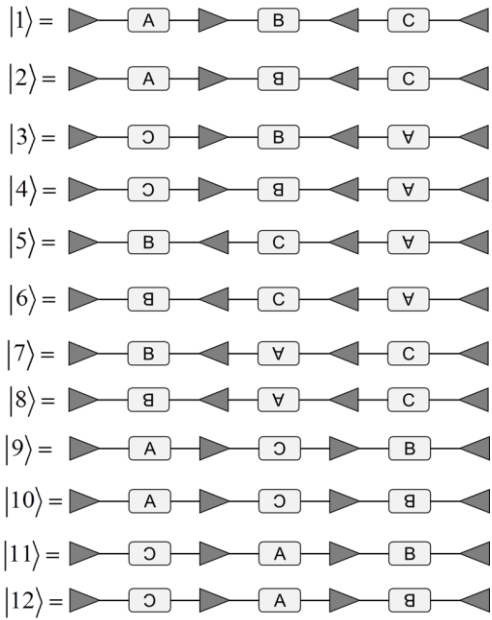

**Figure 12.** For the given initial DNA sequence $|1\rangle$, there are 12 DNA configurations which can be generated by applying inversions on inverted SSR-sites.

We also show (Theorem 3) that after a large number of inversions, when equilibrium distribution of sequences is reached, all of the sequences that can be obtained, are represented with equal probability. This implies that various possible sequences become equally likely after a large number of inversions. To illustrate this here, we simulated the random inversions, according to the assumptions of the Introduction, beginning from the initial configuration $|1\rangle$ (Figure 13A). We find that all 12 configurations listed in Figure 12 appear with the same probability (Figure 13B) as predicted by Theorem 3.

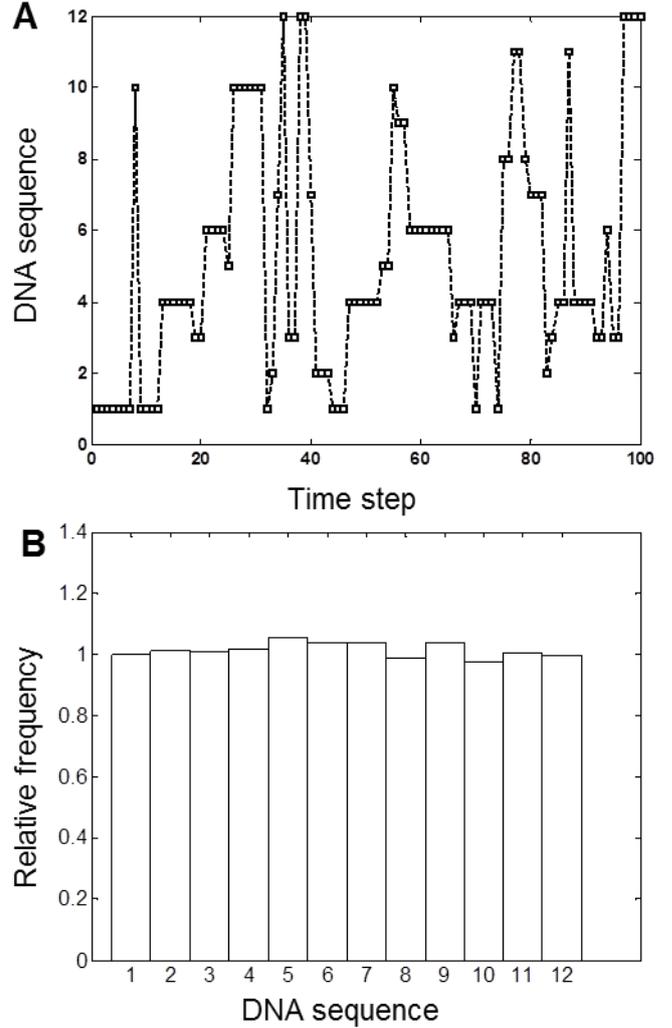

**Figure 13.** (A) Simulation of the first 100 random inversions on initial DNA sequence $|1\rangle$ in Figure 12. Indices of vertical axis represent DNA sequences defined in Figure 12. The probability of an inversion to occur during one step is $r_{inv} = 0.1$. (B) Comparison of frequencies of all 12 possible DNA configurations appearing in 100,000 sequential random inversions. The total number that $|1\rangle$ appears is normalized to 1.

**Example of shufflon sequence in plasmid R64.**

Here we will give an example of applying our results to the bacterial plasmid R64 that contains shufflons (Figure 14). The sequence contains seven SSR-sites and six segments between them. Out of these six segments, only four contain coding regions: A, B, C, and D (Figure 14A). The remaining two segments, denoted by $I_1$ and $I_2$ are non-coding. Here we will calculate the number of combinations possible in the coding regions, using our equations.

Because R64 contains M=1 LL or RR units and N=5 LR

or RL units. The number of combinations of SSR-sites is given by $d_{(1,5)} = C_{M+N}^{N} = C_6^5 = 6$. All of these six SSR-site configurations are shown in Figure 14B. The total number of configurations is given by Eq. (2.2) that yields $Z_{1,5} = 2304$. This number of combinations includes the variance in the non-coding regions $I_1$ and $I_2$. If we are not to include permutations with non-coding regions, we would have to divide this number by 8, the total number of non-coding combinations. Thus get 288 permutations in the coding regions only.

In certain systems, only inversions between RL SSR-sites, but not LR sites are possible (*14*). This case is addressed in Appendix. In the appendix we show that the number of combinations in the constrained case is generally one half of that in the unconstrained case. We thus expect to have 1152 combinations that include both coding and non-coding units. To obtain the number of coding combinations only we have to divide this number by four possible in the constrained case configurations of the non-coding units (see Appendix). We thus obtain 288 combinations for coding sequences only. This number is the same for both unconstrained and constrained inversions. To confirm these numbers in Figures 14B and C we demonstrate how all of these 288 configurations can be obtained with RL inversions.

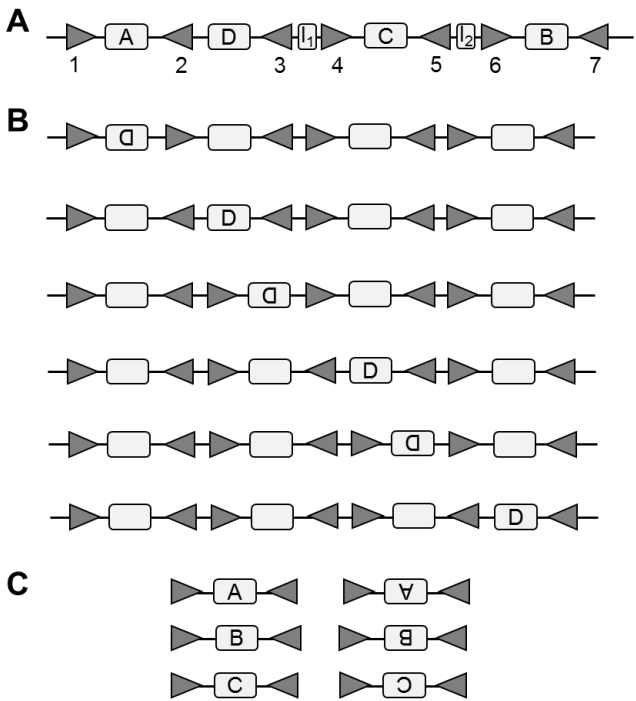

**Figure 14.** (A) The initial DNA sequence with shufflon segments A-D and SSR sites that are called sfx sequences 1-7 (*11*). (B) All possible DNA sequences that can be achieved by the constrained inversions on the initial sequence in (A). Each empty space in sequences in B is filled with one unit from (C), in all possible orders. The total number of possible sequences is $6 \cdot 3! \cdot 8 = 288$.

**Acknowledgement.** The authors thank Tony Zador for suggesting this problem to us and multiple valuable comments. The authors thank Dawen Cai, Jeff Lichtman and Teruya Komano for a helpful communications. This work was supported by NIH R01EY018068 and R01MH092928 and Swartz Foundation. AK acknowledges the hospitality of the Aspen Center for Physics, which is supported in part by NSF Grant No. PHY-1066293.

**Appendix**

**Constrained inversions**

In this section we generalize the results obtained in Sections 2 and 3 to systems in which inversions can happen between inverted SSR-sites ▶◀ (RL) but not between matching SSR-sites ◀▶ (LR) (*14*). We call this type of recombination, when only one type of inversions is possible, the case of constrained inversions. As before, we assume the DNA sequence starts with an R and ends with an L SSR-site. Here we will show that most of the results obtained in the present study can apply in the case of constrained inversions. However, some of sequences cannot be obtained due to the constraint, as detailed below.

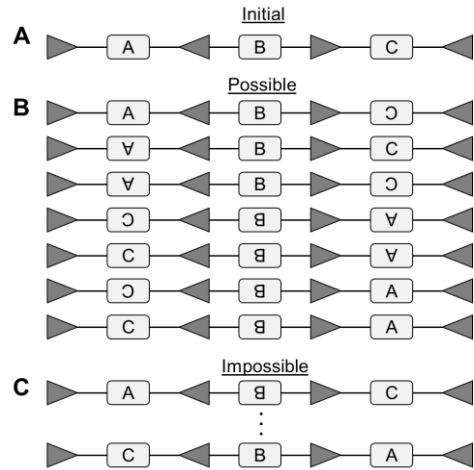

**Figure 15.** Constrained inversions. (A) An initial sequence. (B) Sequences that can be obtained from (A) using inversions between RL but not LR sites. (C) Sequences that cannot be obtained from (A) with the constrained inversions.

Before we present our results, we will illustrate the effects of the constraint on a simple example (Figure 15). This sequence has M=0 LL or RR sites, and N=3 LR or RL sites. Eq. (2.2) yields $Z_{M,N} = 2^N M! [\frac{N+1}{2}]! [\frac{N}{2}]! d_{(M,N)} = 16$ sequences. Here, as above, [x] means the largest integer smaller than or equal to x. However, as illustrated in Figure 15, only 8 of these sequences can be reached using the constrained inversions. We show here that this result is general, i.e. with the constrained inversion, the number of possible sequences in always equal to one half of that for the case of all inversions possible:

$$Z_{M,N}^{constrained} = 2^{N-1} M! [\frac{N+1}{2}]! [\frac{N}{2}]! d_{(M,N)}. \quad (A1)$$

Here $d_{(M,N)}$ is given in Eq. (2.1).

Below we will sketch the proof of Eq. (A1). Let us consider a DNA sequence that includes two LR units. It can be written as follows: >A<B>C<D>E<. Here <B> and <D> are LR, while other units can contain arbitrary combinations of units as well. <B> and <D> cannot be inverted individually without the affecting rest of the sequence. It is easy to check, by enumeration of all possible inversions, that impossible combinations satisfy a simple constraint. Let us introduce the number of reverse-compliments amongst LR units w.r.t. the initial orientation, $t$. Thus, for the sequence >A<D'>C<B>E< (<D'> means reverse-complement of <D>), $t=1$. Let us also introduce the number s, which is the number of exchanges in the <B> and <D> pair. For the sequence >A<D'>C<B>E<, s=1, while for >A<B'>C<D'>E<, s=0. It is possible to check that only the sequences for which s+t is even can be obtained from the initial sequence >A<B>C<D>E<. The sequences for which s+t is odd are not possible through the constrained inversions. This is only true for the fixed remainder of the sequence, i.e. >A<*>C<*>E<. Here '*' denotes either B or D or their reverse-complement. We therefore call the number $\chi=s+t$ the index of sequence. One can obtain >A<*>C<*>E< from >A<B>C<D>E< if $\chi(>A<*>C<*>E<)$ is even.

Let us now consider the sequence with more than two LR elements >A<B>C<D>E<...>Z<. A set of sequences within LR units can be obtained by a permutation P of the original sequence. Permutations form a group of transformations called the symmetric group. Every permutation can be written as a product of several neighboring exchanges. The permutation is called even or odd if it can be written as a product of an even/odd number of neighboring exchanges. Even/odd permutations will be assigned index s equal to 0 or 1, respectively. Although there are several ways to implement P as a superposition of neighboring exchanges, they all have the same index s. The number of reverse complement LR elements can be defined as above, as well as the index χ=s+t. We showed above that a possible exchange does not change the evenness of index χ. Thus, impossible configurations are such that index χ is odd, because the original configuration has an even index. Because for unconstrained inversions both even and odd χ are possible for the same fixed residual sequence, the number of configurations is reduced by a factor of 2 in the case of constrained inversions. Therefore Eq. (A1) describes the number of configurations in the constrained case. This describes the modifications to Theorem 2.

It is possible to show that in the case of constrained inversions, properties 1-4 and Theorem 1 still hold. In the proof of Theorem 1, the first step remains the same while in the second step, we use the operation I=(c,e)(a,c) shown in Figure 16.

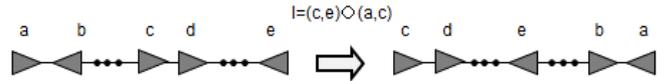

**Figure 16.** Operation involving inversions only between RL SSR-sites that corresponds to Type-B inversion in the proof of Theorem 1

Theorem 3 holds as before, therefore we still have equal probability to observe all possible DNA sequences included in Eq. (A1).